\documentclass[12pt,preprint]{aastex}

\newcommand{\msun} {$M_{\odot}$}
\newcommand{\Te} {$T_{\rm eff}$}

\begin{document}

\title{Stars with Unusual Compositions: Carbon and Oxygen in Cool
  White Dwarf Stars}

\author{P. Dufour
\footnote{D\'epartement de Physique, Universit\'e de
  Montr\'eal, Montr\'eal, QC H3C 3J7, Canada, dufourpa@astro.umontreal.ca}}

\section{INTRODUCTION}

White dwarfs represent the end products of the evolution of all stars
on the main sequence that had initial masses below $\sim$8\msun. After
some important mass-loss episodes in the red giant phases, followed by
the end of thermonuclear activity in the interiors of such stars, they
ultimately shrink to Earth-size objects with masses of about 0.4 to
1.2 \msun. The vast majority of them are believed to be composed
mainly (i.e, more than 99\% by mass) of carbon and oxygen, the
products of hydrogen and helium nuclear burning. Intuitively, then,
one might assume that it would not be surprising to observe a
significant amount of carbon and oxygen in white dwarf
photospheres. However, nuclear burning and mass loss episodes do not
consume 100\% of the hydrogen and helium initially present in each
star at birth. With a surface gravity of the order of log $g\sim$8,
gravitational settling in white dwarf stars is quite efficient and the
light elements leftover from previous evolutionary phases rapidly
float to the surface, while heavier elements sink out of sight. Since
there is ultimately more than enough hydrogen and helium to form an
optically thick photosphere, it is not possible to directly observe
the white dwarf core material. Thus, the majority of white dwarfs are
thus found to have a surface composition that is completely pure in
hydrogen or helium\footnote{Recent studies have revealed that many
  cool white dwarfs also have CaII H \& K lines when observed at
  sufficiently high resolution (see Zuckerman et al., 2010, and
  references therein). The presence of heavy elements in these objects
  is now believed to be the result of the accretion of nearby
  planetesimals or asteroids. Since these objects represent a class of
  their own, and are detailed elsewhere in this volume, this chapter
  contains no further discussion of the impure atmospheres found in as
  DAZ, DZ, and DBZ white dwarfs.}.

Nevertheless, carbon and oxygen features are still found in the
spectra of several hydrogen-deficient objects, namely, the PG~1159
stars, the DQ, DBQ and Hot DQ white dwarfs. In the case of the hot
PG~1159 stars (\Te $\sim$75,000~K and up), the presence of these
elements is somewhat easier to understand since the gravitational
separation of the elements is simply not completed yet. A thorough
review of the observed properties of the extremely hot,
hydrogen-deficient post-asymptotic giant branch (post-AGB) stars has
already been written by Werner and Herwig (2006), and they are not
discussed further here.

In cooler stars, for which the process of radiative levitation can be
considered negligible and gravitational settling has had sufficient
time to transform PG~1159 stars into helium-dominated objects, other
physical mechanisms must be called upon to explain the presence of
observable amounts of carbon (and sometimes oxygen). The following
chapter is a broad review of the observational signatures, physical
properties, and evolution of DQ, DBQ and Hot DQ white dwarfs, and also
present an overview of the main challenges that future investigations
of these types of object should try to address. Although these
spectral types together represent only a small fraction of the total
number of white dwarfs, they nevertheless provide extremely valuable
information about the evolution of stars following the AGB phase as
well as the spectral evolution of white dwarfs in general.

{\bf [The 45 page chapter, complete with figures, tables, and
  references can be found in its entirety in "White Dwarf Atmospheres
  and Circumstellar Environments", ed. D. W. Hoard, Wiley-VCH, ISBN
  978-3-527-41036-6, published in September 2011.]}

\end{document}